\documentclass[reprint,
 amsmath,amssymb,
aps,
pra,
]{revtex4-2}
\usepackage{mathrsfs}
\usepackage{graphicx}
\usepackage{dcolumn}
\usepackage{bm}
\usepackage{float}
\usepackage{subfig}
\usepackage{color}

\begin{document}

\preprint{APS/123-QED}

\title{Analytical Study of the Non-Hermitian Semiclassical Rabi Model}

\author{Yibo Liu}
\affiliation{%
	 Zhejiang Key Laboratory of Micro-Nano Quantum Chips and Quantum Control, School of Physics, Zhejiang University, Hangzhou 310027, China
}
\author{Liwei Duan}
\affiliation{%
	Department of Physics, Zhejiang Normal University, Jinhua 321004, China
}%
\author{Qing-Hu Chen}%
 \email{qhchen@zju.edu.cn}
\affiliation{%
  Zhejiang Key Laboratory of Micro-Nano Quantum Chips and Quantum Control, School of Physics, Zhejiang University, Hangzhou 310027, China \\
 Collaborative Innovation Center of Advanced Microstructures, Nanjing University, Nanjing 210093, China
}%

\date{\today}

\begin{abstract}
The $\mathcal{PT}$ symmetric semiclassical Rabi model explores  the fundamental interaction  between a two-level atom and a classical field, revealing  novel phenomena in open systems through the inclusion  of non-Hermitian terms. We propose a single similarity transformation that yields an effective Hamiltonian in  rotating-wave approximation, enabling an analytical solution. The phase boundary of the $\mathcal{PT}$-broken phase, derived from the analytical eigenvalues, closely matches  the numerical  exact one over a wide range of atomic frequencies, demonstrating the effectiveness of the analytical approach, especially  at the main resonance. The Floquet parity operator is also introduced,  providing a deeper physical understanding of the emergence of the $\mathcal{PT}$-broken phase. Furthermore, by analyzing the dynamics of excited-state population, we observe  several stable oscillations in the Fourier spectrum, demonstrating the applicability of the  analytical method beyond the single-photon resonance region. The Bloch-Siegert shift is also discussed and,  surprisingly,  resembles its Hermitian counterpart, except for the higher-order terms in the coupling strength.  The present analytical treatment  provides a concise and accurate description  of the main physics of this non-Hermitian atom-field interaction system.
\end{abstract}

\maketitle

\section{introduction}

{The Rabi model, describing the interaction between a
two-level system and a classical driving field, is a cornerstone of quantum
optics and atomic physics
~\cite{Rabi1936,Rabi1937,Shirley1965,Braak2016,gerry2023}. In recent years,
there has been growing interest in extending the Rabi model to non-Hermitian
systems, where the Hamiltonian is no longer self-adjoint
~\cite{Joglekar2014,Lee2015,Gong2015,Xie2018,Quijandria2018,Wang2021,Lu2023}. The non-Hermitian systems, usually accompanied with complex eigenvalues, can exhibit rich and novel phenomena such as non-Hermitian phase transition ~\cite{wei2017,Longhi2019,Fahri2021}, non-Hermitian quantum Hall effect ~\cite{yoshida2019,ochkan2024}, non-reciprocal dynamics \cite{reisenbauer2024,Brighi2024}, non-Hermitian quantum decoherence ~\cite{Castagnino2012,dey2019}, and so on. The non-Hermitian Hamiltonian provides a framework for investigating open systems and facilitates a deep understanding of the phenomena associated with non-Hermitian physics ~\cite{Yuto2020}.}

{In particular, one of the most intriguing features of
non-Hermitian systems is the concept of parity-time ($\mathcal{PT}$)
symmetry, where the combined operation of parity (spatial inversion) and
time-reversal symmetry allows for real eigenvalues even in the absence of
Hermiticity
~\cite{Leggett1987,Bender1998,Konotop2016,El-Ganainy2018,bender2023}. When the system parameters cross certain critical values, the $\mathcal{PT}$ symmetry is broken, leading to the emergence of complex eigenvalues and the formation of exceptional points (EPs), where the eigenvalues and eigenstates coalesce ~\cite{moiseyev2011}. The $\mathcal{PT}$-symmetric non-Hermitian systems display numerous novel phenomena, such as $\mathcal{PT}$-symmetric phase transition ~\cite{bender2013,Konotop2016}, sensitivity enhancement ~\cite{Liu2016,Li2016,Farhat2020}, topological insulators ~\cite{Hu2011,Ni2018,Ezawa2021,fritzsche2024}, and it thus offers some new physics which is absent in the Hermitian systems ~\cite{Guo2009,feng2017,Kawabata2019}.}

{Although the time-independent non-Hermitian systems has
been well developed, the question remains challenging when extended to
time-dependent situations ~\cite{Fring2017}. Several recent works have
focused on time evolution and have uncovered a big amount of intriguing dynamical behaviors
~\cite{Luo2013,Lee2015,Gong2015,doppler2016,Hassan2017,Xie2018,Koutserimpas2018,Zhang2019,Wang2021}. Some efforts have been made to investigate the nature of the non-Hermitian Rabi model (NHRM) ~\cite{Lee2015,Gong2015,Xie2018,li2019}. The NHRM can also be regarded as a periodically driven non-Hermitian two-level system ~\cite{Bender1998}. The time-independent non-Hermitian two-level system  ~\cite{Bender1998} has been extensively studied in the literature and can be mapped to many important non-Hermitian systems. Hoverer,  an exact analytical solution to the NHRM remains elusive, and the physical picture remains unclear. In this study, we aim to find a solvable analytical effective Hamiltonian and gain deeper understanding of the $\mathcal{PT}$-symmetric NHRM.}

The paper is structured as follows: In Sec. II, by using a
similarity transformation, we transform the NHRM into the rotating-wave
approximation (RWA) form with renormalized parameters, making the NHRM
effectively solvable. In Sec. III, the analytical $\mathcal{PT}$ spectrum
is obtained analytically from the  effective solvable model. A comparison
with the numerically exact results is performed.  We introduce the Floquet parity operator and explain how the $\mathcal{PT}$-broken phase emerges in Sec. IV. Section V is devoted
to an analytical analysis of dynamics. The Fourier spectrum analysis is also presented in this section. The analytical Bloch-Siegert shift is  presented and analyzed in Sec. VI. The last section contains some concluding remarks.

\section{Model and formalism}

The Hamiltonian of the NHRM with the purely imaginary coupling strength ~\cite%
{Lee2015} can be expressed as
\begin{equation}
\hat{H}(t)=\frac{\Delta }{2}\hat{\sigma}_{z}+i\frac{A}{2}\cos \omega t\hat{%
\sigma}_{x},  \label{Hamiltonian}
\end{equation}%
where $\Delta $ represents the atomic frequency, $A$ denotes the coupling
strength between the atom and the external field, and $\omega $ stands for
the field frequency. The operators $\hat{\sigma}_{k}$ represent the
components of the Pauli matrix, where $k=x,y,z$. Notably, this Hamiltonian
exhibits $\mathcal{PT}$ symmetry, which can be demonstrated as
\[
\lbrack \mathcal{PT},\hat{H}(t)]=0,
\]%
where the parity operator $\mathcal{P}=\hat{\sigma}_{z}$ acts as $\hat{\sigma%
}_{x}\rightarrow -\hat{\sigma}_{x}$, and the time-reversal operator $%
\mathcal{T}$ transforms $i\rightarrow -i$ and $t\rightarrow -t$. In the $%
\mathcal{PT}$-unbroken phase, all quasi-energies are real, whereas in the $%
\mathcal{PT}$-broken phase, the quasi-energies become complex.

In recent experiments, optical and radio-frequency fields provide an effective tool for manipulating atom loss, enabling the implementation of the NHRM Hamiltonian through time-periodic coupling between two spins in a dissipative system of ultracold atoms~\cite{li2019}.	The two lowest $^2S_{1/2}$ hyperfine levels in a non-interacting Fermi gas of $^6$Li atoms are prepared.	A resonant optical beam excites the atoms from the lower level to the $2P$ excited state, generating atom loss in the lower level.	This state-dependent, purely lossy system allows the lossy Hamiltonian to be mapped onto the $\mathcal{PT}$-symmetric NHRM (\ref{Hamiltonian}).	

The structure of the phase diagram of the NHRM has been theoretically studied using   perturbation theory ~\cite{Shirley1965}, as discussed in Refs. ~\cite{Joglekar2014,Lee2015}.	For the Hermitian semi-classical Rabi model, an analytically exact solution remains unavailable. However, some analytical studies based on perturbation theory ~\cite{Hausinger2010} and unitary transformations ~\cite{Lv2012,Yan2015,Han2024} have been proposed, in addition to numerical solutions based on Floquet theory. For the NHRM, developing an accurate analytical scheme to bridge the gap between perturbation theory and the numerical exact solution would be valuable.	

To facilitate the analytical study, we transform the above Hamiltonian in Eq.~(%
\ref{Hamiltonian}) into an effective RWA. Similar to the unitary
transformation in the Hermitian semi-classical RM ~\cite{Lv2012,Irish2022},  we introduce a similarity transformation with a generator in the present NHRM,
\begin{equation}
\hat{S}(t)=\frac{A}{2\omega }\sin (\omega t)\alpha \hat{\sigma}_{x},
\label{transfer}
\end{equation}%
where $\alpha $ is a coefficient to be determined later. We can then 
obtain the following transformed Hamiltonian
\begin{align}
\hat{H}^{\prime }(t)& =e^{\hat{S}(t)}\hat{H}(t)e^{-\hat{S}(t)}-ie^{\hat{S}%
(t)}\left( \frac{\partial }{\partial t}e^{-\hat{S}(t)}\right)  \nonumber \\
& =\frac{\Delta }{2}\left\{ \cos \left[ i\frac{A}{\omega }\sin (\omega
t)\alpha \right] \sigma _{z}+\sin \left[ i\frac{A}{\omega }\sin (\omega
t)\alpha \right] \sigma _{y}\right\}  \nonumber \\
& +\frac{iA}{2}\left( 1-\alpha \right) \cos \left( \omega t\right) \sigma
_{x}.
\end{align}%
$\hat{H}^{\prime }(t)$ can be divided into three
parts $\hat{H}^{\prime }(t)=H_{0}^{\prime }+H_{1}^{\prime }+H_{2}^{\prime }\
$
\begin{align}
H_{0}^{\prime }& =\frac{\Delta }{2}J_{0}\left( i\frac{A}{\omega }\alpha
\right) \sigma _{z}, \\
H_{1}^{\prime }& =\Delta J_{1}(i\frac{A}{\omega }\alpha )\sin (\omega
t)\sigma _{y}+\frac{iA}{2}\left( 1-\alpha \right) \cos \left( \omega
t\right) \sigma _{x}, \\
H_{2}^{\prime }& =\Delta \sum_{n=1}^{\infty }J_{2n}\left( i\frac{A}{\omega }%
\alpha \right) \cos (2n\omega t)\sigma _{z}  \nonumber \\
& +\Delta \sum_{n=1}^{\infty }J_{2n+1}\left( i\frac{A}{\omega }\alpha
\right) \sin \left[ (2n+1)\omega t\right] \sigma _{y},
\end{align}%
where $J_{n}(x)$ is the $n$th-order Bessel function of the first kind, and
the following identities have been used
\begin{align}
\cos (\gamma \sin t)& =J_{0}(\gamma )+2\sum_{n=1}^{\infty }J_{2n}(\gamma
)\cos 2nt, \\
\sin (\gamma \sin t)& =2\sum_{n=0}^{\infty }J_{2n+1}(\gamma )\sin (2n+1)t.
\end{align}

Note that $H_{0}$ is the time-independent component, while $H_{1}$
oscillates with a single-harmonic frequency. The term $H_{2}^{\prime }$
includes all higher-order harmonic terms and Bessel functions with orders $%
n\geq 2$. By neglecting $H_{2}^{\prime }$, we can derive the effective
Hamiltonian that retains only a single-harmonic frequency, which is $%
H^{\prime }\simeq H_{0}^{\prime }+H_{1}^{\prime }$.

At this point, the coefficient $\alpha $ can be determined by solving the
equation,
\begin{equation}
\Delta I_{1}(\frac{A}{\omega }\alpha )=\frac{A}{2}\left( 1-\alpha \right) ,
\label{transition}
\end{equation}%
where $I_{n}(x)$ is $n$th-order modified Bessel function, and  the identity $J_{n}(ix)=i^{n}I_{n}(x)$  has been used. The transformed Hamiltonian
now takes an RWA-type form with a non-Hermitian coupling strength,
\[
H^{\prime }(t)=\frac{\Delta }{2}I_{0}\left( \frac{A}{\omega }\alpha \right)
\sigma _{z}+i\Delta I_{1}(\frac{A}{\omega }\alpha )\left( e^{-i\omega t}\hat{%
\sigma}_{+}+e^{i\omega t}\hat{\sigma}_{-}\right) .
\]%
Applying a unitary transformation $\hat{R}(t)=\exp \left( i\omega t\hat{%
\sigma}_{z}/2\right) $, we obtain a time-independent Hamiltonian as
\begin{equation}
\tilde{H}=\frac{\tilde{\Delta}}{2}\hat{\sigma}_{z}+i\frac{\tilde{A}}{4}\hat{%
\sigma}_{x}.  \label{tiemh}
\end{equation}%
where
\begin{eqnarray*}
\tilde{\Delta} &=&\Delta I_{0}\left( \frac{A}{\omega }\alpha \right) -\omega
, \\
\tilde{A} &\equiv &2A\left( 1-\alpha \right) .
\end{eqnarray*}%
Here, the renormalized atom characteristic frequency $\tilde{\Delta}$ can
be interpreted as an effective detuning, while the effective coupling
strength $\tilde{A}$ is given by by Eq. ~(\ref{transition}). Thus, by
performing a single similarity transformation and neglecting 
higher-order harmonic terms, we derive an effective non-Hermitian two-level
atomic Hamiltonian, which forms the foundation of our subsequent
work. 

\section{$\mathcal{PT}$ phase diagram and quasi-eigenvalues spectrum}

\begin{figure}[t]
\includegraphics[width=\linewidth]{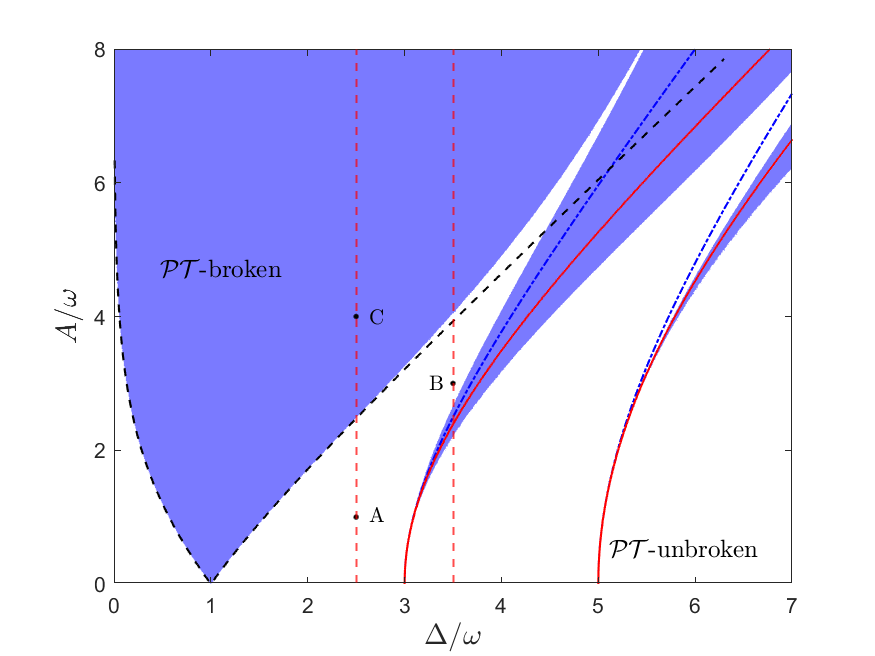}
\caption{$\mathcal{PT}$ phase diagram. The numerical $\mathcal{PT}$-broken phases marked in blue and the $\mathcal{PT}$-unbroken phases  in white. The black dashed line denotes the EPs from the present analytical scheme,  as given by by Eq.~(\ref{quasienergies}). Higher-order $\mathcal{PT}$-broken phase regions are predicted by analytical method (red solid line) and perturbation theory (blue dash-dotted line) ~\cite{Lee2015}. The quasi-eigenvalue spectrum along the two vertical red dashed lines will be shown in Fig.~\protect\ref%
{fig:spectrum}, and the dynamics at points A, B, and C will be presented in Fig.~\protect\ref{fig:dynamics}. }
\label{fig:ptphase}
\end{figure}

\begin{figure}[t]
\subfloat{\includegraphics[width=\linewidth]{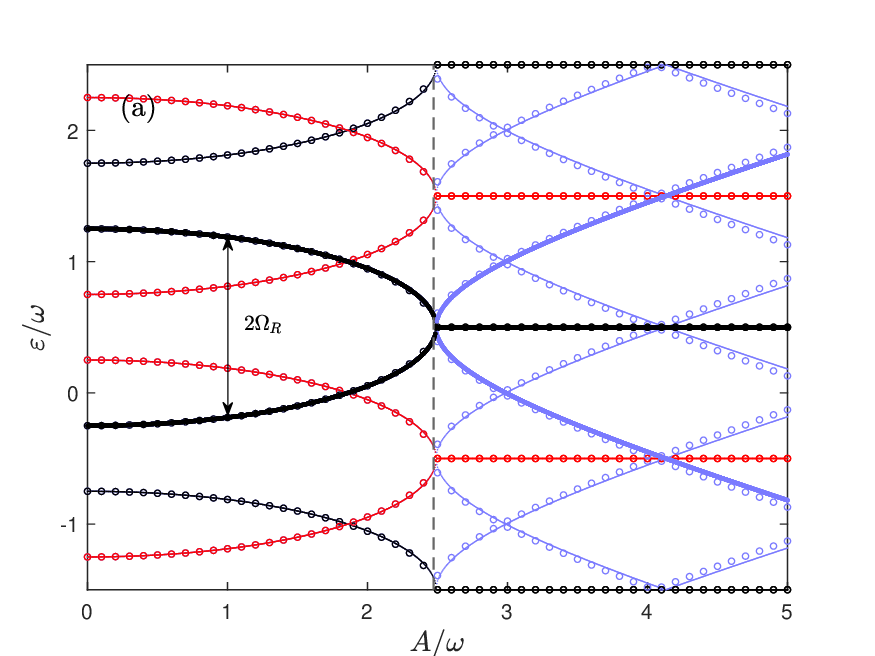}}\vspace{-2pt}
\subfloat{\includegraphics[width=\linewidth]{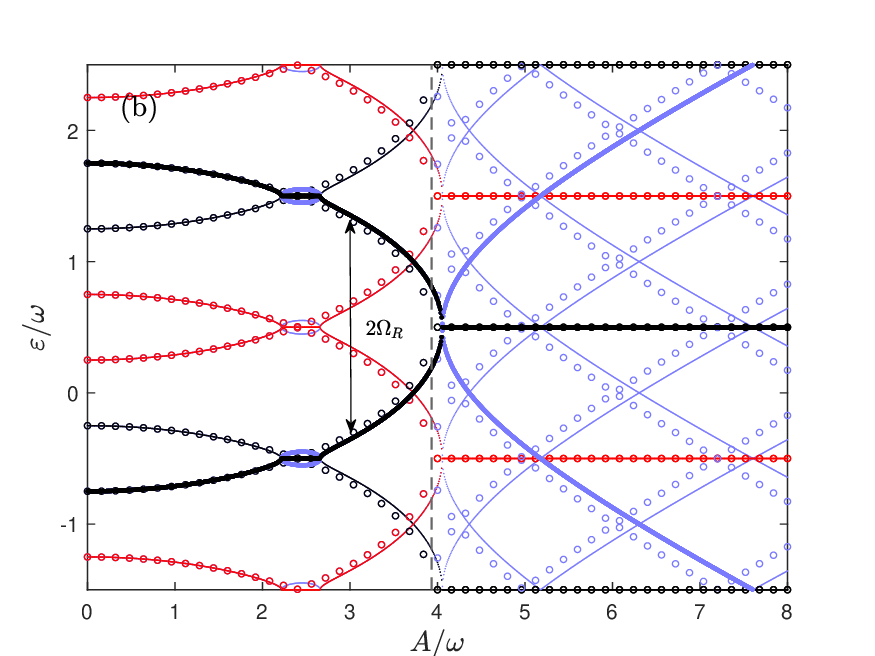}}
\caption{The quasi-eigenvalue spectrum $\varepsilon/\omega$ as a function of the coupling strength $A/\protect\omega$ for (a) $\Delta/\protect\omega = 2.5$  and (b) $\Delta/\protect\omega = 3.5$. The black (red) lines represent the real part of the numerical  quasi-eigenvalues  in the odd (even) parity, and the violet lines denote their
imaginary parts. The open circles represent the  corresponding  analytical results. The  real part of the numerical quasi-eigenvalues for $n=0$ is marked by thick black lines.  The EPs from analytical quasi-eigenvalue equation (\ref{quasienergies}) are shown by vertical dashed line. The analytical  Rabi frequencies are indicated by the double arrows, which will shown in Eq. ~(\ref{PE}) and Fig. ~\ref{fig:fs}. }
\label{fig:spectrum}
\end{figure}

An essential characteristic of $\mathcal{PT}$-symmetric models is their
spectrum. Despite the non-Hermitian nature of the Hamiltonian,  all eigenvalues become real under certain parameter conditions. We designate the system as being in a $\mathcal{PT}$-unbroken phase
when these parameters guarantee real eigenvalues. Conversely, when the
parameters result in the appearance of imaginary parts in the eigenvalues, we
characterize the system as being in a $\mathcal{PT}$-broken phase.

In the time-periodic model, the eigenvalues of the instantaneous Hamiltonian
do not have the same physical meaning as those in time-independent
scenarios. Therefore, we must study the quasi-eigenvalues. Fortunately, we derived a solvable effective time-independent Hamiltonian in Eq.~(\ref{tiemh}) in the previous section, allowing us to readily obtain its eigenvalues.	

\begin{equation}
\varepsilon _{\pm }=\pm \frac{1}{2}\sqrt{\tilde{\Delta}^{2}-\frac{\tilde{A}%
^{2}}{4}}+\frac{\omega }{2}.  \label{quasienergies}
\end{equation}
The exceptional points are given by $\tilde{\Delta}^{2}=\tilde{A}%
^{2}/4 $, as derived from Eq.~(\ref{quasienergies}), which directly yield the $\mathcal{PT}$
phase diagram directly in Fig.~\ref{fig:ptphase}. When $\tilde{\Delta}^{2}>
\tilde{A}^{2}/4$, the model is in a $\mathcal{PT}$-unbroken phase,
characterized by real eigenvalues.  Conversely, when  $\tilde{\Delta}^{2}<\tilde{A}%
^{2}/4$, the eigenvalues  become complex and conjugate in pairs. The black
dashed line in Fig.~\ref{fig:ptphase} separates the two phases, which are
just the EPs.

To solve the time-periodic NHRM numerically, we apply Floquet theory to
obtain the quasi-eigenvalues. For the NHRM Hamiltonian (\ref%
{Hamiltonian}), the Floquet Hamiltonian is given by $\hat{\mathcal{F}}=\hat{H%
}(t)-i\frac{\partial }{\partial t}$. As detailed in  Appendix A, we perform exact
diagonalization (ED)  on the matrix form of the Floquet
Hamiltonian. Using the numerically obtained quasi-eigenvalues, we determine
the boundary between the $\mathcal{PT}$-broken and $\mathcal{PT}$-unbroken phases, as shown in Fig.~\ref{fig:ptphase}, marked by the blue-shaded region. 

Interestingly, the present analytical phase boundary (black dashed lines) agrees excellently with with the exact numerical results over a wide atomic frequency range, up to $\Delta/\omega \approx 3$ (the three-photon resonance), which constitutes the primary region for the $\mathcal{PT}$-broken phase. When $\Delta /\omega >3$, additional $\mathcal{PT}$-broken
phases intervene the $\mathcal{PT}$-unbroken region. These secondary $%
\mathcal{PT}$-broken phases, however, are not captured by the current
analytical scheme.  Is the present analytical scheme unable to provide
any information about the secondary $\mathcal{PT}$-broken phases? The answer
is "no".

We present the eigenvalues from both the analytical scheme ( Eq.~(\ref{quasienergies}), open circles) and the Floquet theory (thick curves) in
Fig.~\ref{fig:spectrum} for $\Delta /\omega =2.5$ and $3.5$.  These two
values are marked by two vertical red dashed lines in Fig.~\ref%
{fig:ptphase}, both cross the primary $\mathcal{PT}$-broken phase boundary,
but $\Delta /\omega =3.5$ line also crosses the secondary $\mathcal{PT}$%
-broken phase boundary. According to the Floquet theorem, for a
time-periodic Hamiltonian, the quasi-eigenvalues shifted by $n\omega $ ($n$
is an integer) do not change the physical state, similar to a generalized
Brillouin zone in the Bloch band theory.  For later reference, we also display some shifted
quasi-eigenvalues for both the analytical scheme (Eq. ~(\ref{quasienergies})) and
Floquet theory.

Figure~\ref{fig:spectrum} (a) shows that the analytical results closely agree 
with the numerical quasi-energies for the coupling strengths up to $A\simeq
6\omega $ for $\Delta /\omega =2.5$. The analytical EPs, indicated by dashed
lines, are nearly identical to the numerical ones.

As shown in Fig.~\ref{fig:spectrum} (b) for$\ \Delta /\omega =3.5$, the
analytical quasi-eigenvalues agree with the numerical results, except in
the small, ellipse-shaped region around quasi-eigenvalues $\varepsilon
/\omega =(2n+1)/2$. In this region, the analytical scheme can only provide
real quasi-eigenvalues, and level crossings occur, while the numerical
results yield complex quasi-eigenvalues. Further analysis will be presented
in the next section to explain these observations, based on the uncovered
symmetry of the NHRM within the framework of Floquet theory.

\section{Floquet parity operator and spectrum analysis}

To explain the $\mathcal{PT}$-symmetry breaking and the EPs discussed in the previous section, we introduce the Floquet parity operator
\[
\hat{\Pi}=e^{i\pi (\hat{\sigma}_{+}\hat{\sigma}_{-}+\hat{G})}=-\hat{\sigma}%
_{z}(-1)^{\hat{G}},
\]%
where $\hat{G}=\sum_{n=-\infty }^{\infty }n|n\rangle \langle n|$ is  constructed using
the state vector $|n\rangle $ from an orthonormal,
complete Fourier basis set $\exp \left( -in\omega t\right) $ in the Hilbert
space of   $2\pi /\omega $-periodic function. It can be shown that $\hat{\Pi%
}\mathcal{F}\hat{\Pi}^{\dagger }=\mathcal{F}$, where $\mathcal{F}=\hat{H}%
-i\partial /\partial t$ is  the Floquet Hamiltonian. This implies that
the Hilbert space of the Floquet Hamiltonian can be decomposed into two
independent subspaces, corresponding to even and odd Floquet parity, with
eigenvalues $\Pi =\pm 1$, respectively.

For odd (even) parity ($\Pi =\mp 1$), the subspace basis should be $\{\cdots
|\mp ,-1\rangle ,|\pm ,0\rangle ,|\mp ,1\rangle ,\cdots \}$ , so that the
Floquet Hamiltonian can be expressed as
\begin{equation}
\begin{pmatrix} \ddots & & & & & \\ & \mp\frac{\Delta}{2}-\omega &
i\frac{A}{4} & 0 & 0 & \\ & i\frac{A}{4} & \pm\frac{\Delta}{2} &
i\frac{A}{4} & 0 & \\ & 0 & i\frac{A}{4} & \mp\frac{\Delta}{2}+\omega &
i\frac{A}{4} & \\ & 0 & 0 & i\frac{A}{4} & \pm\frac{\Delta}{2}+2\omega & \\
& & & & & \ddots\end{pmatrix}. \label{matrix_f}
\end{equation}
We can also numerically diagonalize  this matrix for both even and odd parity to
calculate the corresponding quasi-eigenvalues.

The quasi-eigenvalues, denoted by the thick curves in Fig.~\ref{fig:spectrum},
correspond to the odd parity subspace. Quasi-eigenvalues shifted by $%
2n\omega $, marked in black, also belong to this subspace. In contrast, the
quasi-eigenvalues of even parity are shifted by $(2n+1)\omega $, as marked in
red. In principle, quasi-eigenvalues with different parities may cross.
However, within the same parity subspace, they cannot intersect without an
additional symmetry. Fortunately the present NHRM just possesses an
alternative $\mathcal{PT}$ symmetry, which enables quasi-eigenvalues within the
same parity subspace to intersect, leading to EPs. This phenomenon underlies
the $\mathcal{PT}$ symmetry breaking and the emergence of EPs.

The  picture above illustrating the $\mathcal{PT}$ symmetry breaking 
fully describes the features of the quasi-eigenvalue spectrum shown in Fig.~\ref{fig:spectrum}. Two branches of the thick curve (representing  real values), both with the
same odd parity, approach each other as  $A/\omega$ increases and  intersect at the vertical dashed line, forming
the primary EPs. These EPs can be accurately  captured by $\tilde{\Delta}^{2}=%
\tilde{A}^{2}/4$ in Eq.~(\ref{quasienergies}).

As mentioned above,  the eigenvalue $\varepsilon _{-}$, shifted by $2n\omega $, retains the  same parity. If the shifted $\varepsilon _{-}$ cross with $\varepsilon_{+}$, i.e., $\varepsilon _{+}=\varepsilon _{-}+2n\omega$, then  we have
\begin{equation}
\sqrt{\tilde{\Delta}^{2}-\frac{\tilde{A}^{2}}{4}}=2n\omega.   \label{cross}
\end{equation}
This equation gives the coupling strength $A/\omega $ for a given $\Delta /\omega $, at which the real eigenvalues cross, thus corresponding to the $\mathcal{PT}$-broken region. The primary EP   corresponds to $n=0$ in this equation. We can also determine the crossing points where real eigenvalues with the same parity intersect, for any $n > 0$.	

The lines representing the crossing points for $n=1$ and $n=2$ are shown as red solid lines in Fig.~\ref{fig:ptphase}, both lying within the $\mathcal{PT}$-broken region.	Using perturbation theory, Lee et al. predicted a line of maximal $\mathcal{PT}$ symmetry breaking (where the imaginary eigenvalue is maximized) in Ref.~\cite{Lee2015},  shown as a blue dash-dotted line in Fig.\ref{fig:ptphase} for comparison.	The two lines agree well for small $ A/\omega $ but begin to deviate as  $A/\omega $ increases. Noticeably, their line for $n=2$ does not lie within the $\mathcal{PT}$-broken region, while ours does.	Therefore, our analytical scheme can more accurately predict the onset of $\mathcal{PT}$ symmetry breaking, beyond the primary $\mathcal{PT}$-broken phase, particularly when $\Delta$ exceeds $5\omega$.	

If $A$ is very small, and higher-order terms in $A$ are neglected, we can derive a closed-form solution to Eq.~(\ref{cross}),
\begin{equation}
    A=\frac{\Delta+\omega}{\Delta}\sqrt{(\Delta-\omega)^2-4n^2\omega^2}.\label{small}
\end{equation}
Based on the similarity transformation in Sec. II, this result is exact in the limit as $A/\omega \to 0$. Since $\Delta = (2n + 1) \omega$, the crossing point occurs at $A/\omega = 0$. This indicates that the $\mathcal{PT}$-broken line  intersect the $A/\omega = 0$ axis at these resonance points.	

As shown in Fig.~\ref{fig:ptphase} and in Ref.~\cite{Lee2015}, the second and third $\mathcal{PT}$-broken phases, for $n=1$ and $2$, are expected to be extremely narrow at small values of $A/\omega$ near 0, but they do not extend to the $A/\omega = 0$ axis according to numerical calculations.	This discrepancy arises from the numerical calculations, which require truncation, while  the exact solution can only be obtained using an
infinite basis. At the single-photon resonance $\Delta = \omega $ for $n=0$,  Eq.~(\ref{small}) gives $A/\omega = 0$, consistent with the primary $\mathcal{PT}$-broken region, which touches $A/\omega = 0$ axis at $\Delta/\omega = 1$, as shown in Fig.~\ref{fig:ptphase}.

We now focus on the more complicated Floquet quasi-eigenvalues spectrum in
Fig.~\ref{fig:spectrum} (b) for $\Delta= 3.5\omega $. Specifically, within the
small, ellipse-shaped region around quasi-eigenvalues $\varepsilon
=(2n+1)\omega /2$, the present analytical results from Eq.~(\ref{quasienergies}),
denoted by the open circles, show real level crossings. The crossing point at $A/\omega=2.39$ can be calculated  from Eq. ~(\ref{cross}) for  $n=1$. This is also the point where the first red solid line intersects the second vertical red dashed line in Fig.\ref{fig:ptphase}.  However, the two real quasi-eigenvalues with the same parity cannot cross. As a result, an
imaginary part emerges to resolve this issue, signaling the new PT symmetry
breaking. The analytical results in Eq.~(\ref{quasienergies}) cannot fully
describe, but can predict this $\mathcal{PT}$-broken region. In contrast, 
real spectrum lines with different parities can cross at $\varepsilon =n\omega
$.

When $\Delta /\omega =(2n+1)$, i.e. $(2n+1)$
-photon resonance, the two quasi-eigenvalues with the same parity coincide at $A/\omega=0$, as  proven in Eq.~ (\ref{small}) for any $n$.
If $\Delta $ increases slightly, the two quasi-eigenvalues will
intersect. However, they cannot cross because they share the same parity. As
a result, a complex quasi-eigenvalue must emerge to prevent the crossing of
real quasi-eigenvalues, leading to the formation of EP points.

If $3<\Delta /\omega <5$, as demonstrated in Fig.~\ref{fig:spectrum} (b),
the two adjacent quasi-eigenvalues obtained from the numerical diagonalization of the matrix (\ref{matrix_f}), which share the same parity (both black or both red lines), should avoid crossing each other at a finite value of $A/\omega$.	 Consequently, a $\mathcal{PT}$-broken phase emerges, indicated by small, ellipse-shaped blue curves representing the imaginary part of the quasi-eigenvalues. This explains the origin of the second $\mathcal{PT}$-broken region in Fig.~\ref{fig:ptphase}. However, two
adjacent real quasi-eigenvalues from the analytical scheme cross in the
$\mathcal{PT}$-broken region. In other words, the analytical scheme cannot fully describe this
$\mathcal{PT}$-broken region, but can predict its location. If we further increase
$\Delta /\omega $ to the range $5<\Delta /\omega <7$, both adjacent and
next-adjacent numerical quasi-eigenvalues  avoid the real quasi-eigenvalue crossing.
Consequently, two $\mathcal{PT}$-broken phases emerge, as illustrated  in
Fig.~\ref{fig:ptphase} by the second and  third $\mathcal{PT}$-broken
regions. At large {$A/\omega $, the widths of the different }$\mathcal{PT}$%
-broken regions can become large enough to merge into a single, contiguous $%
\mathcal{PT}$-broken region.

So far, we have explained the origin of the $\mathcal{PT}$-broken phases and
accounted for the phase diagram using the conserved
Floquet parity operator. Additionally, the Floquet parity operator can 
be applied to the Hermitian semi-classical Rabi model. Within the same parity
subspace, real quasi-energies cannot cross but instead exhibit an avoided-crossing
feature {~\cite{Shirley1965}}. As shown in Appendix A, quasi-energies with
different parities can cross, whereas quasi-energies with the same parity can
only avoid each other. The present non-Hermitian model exhibits richer and more complex properties than its Hermitian counterpart, due to the additional $\mathcal{PT}$ symmetry.	

Our current analytical scheme, based on a single similarity transformation,
accurately describes  the phase diagram for $\Delta /\omega <3$ and can
also predict the new $\mathcal{PT}$-broken region at $\Delta /\omega >3$. 
By combining the primary $\mathcal{PT}$-broken region for $n=0$, determined by Eq.~(\ref{quasienergies}), with the $\mathcal{PT}$-broken line from Eq.~(\ref{cross}) for $n>0$, as shown in Fig.~\ref{fig:ptphase}, our analytical scheme can produce  a highly accurate, nearly exact phase diagram within $A/\omega < 2$ for arbitrary $\Delta/\omega$, except for the small region of the second $\mathcal{PT}$-broken phase. In the next section, we will examine whether it accurately captures  the dynamics.	

\section{Dynamics of the two-level system}

\begin{figure}[t]
\includegraphics[width=\linewidth]{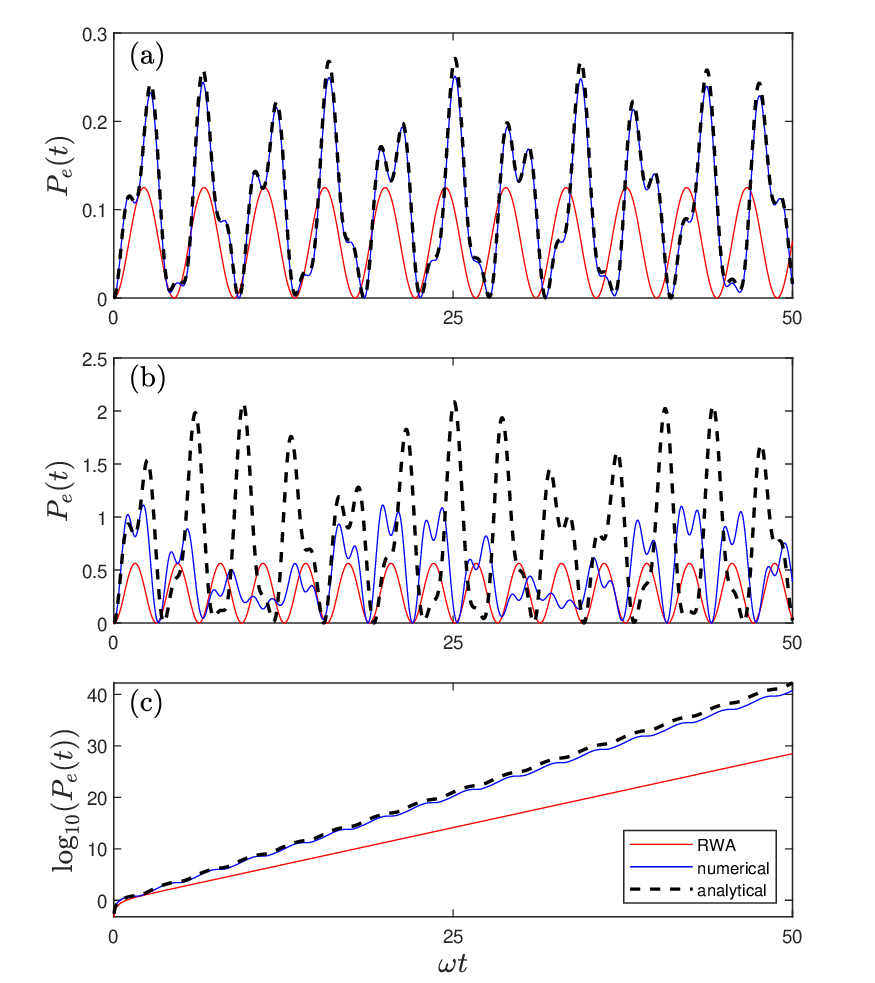}
\caption{Time evolution of the excited state population $P_{e}(t)$ at three  points in the phase diagram shown in Fig.~\ref{fig:ptphase}:  (a) Point A with $\Delta=2.5\protect\omega $, $A=1\protect\omega $; (b) Point B with $%
\Delta=3.5\protect\omega $, $A=3\protect\omega $; and (c) Point C with $\Delta=2.5%
\protect\omega $, $A=4\protect\omega $. The analytical  results from Eq.~ (\ref{PE}) are denoted by black dashed lines, and the numerical  results are represented by the blue solid line. The red line denotes the results in
rotating-wave approximation from Eq.~ (\ref{RWA}) . }
\label{fig:dynamics}
\end{figure}

In this section, We explore the time-evolution of the two-level system. 
The time dependent wavefunction can be represented as $|\psi (t)\rangle
=(c_{+}(t)\ c_{-}(t))^{\text{T}}$, where $c_{+}(t)$ and $c_{-}(t)$ denote
the probability amplitudes of the upper and lower states, respectively, in a
two-level atom. Using the Schr\"{o}dinger equation, we can derive the
following two coupled first-order differential equations:
\begin{align}
i\dot{c}_{+}& =\frac{\Delta }{2}c_{+}+i\frac{A}{2}\cos \omega tc_{-}, \\
i\dot{c}_{-}& =-\frac{\Delta }{2}c_{-}+i\frac{A}{2}\cos \omega tc_{+}.
\end{align}%
The exact dynamics can be determined by numerically solving two coupled
dynamical equations, as an analytical solution is highly challenging. An
effective approximation involves applying the RWA to decouple the
probability amplitudes. This approach simplifies the calculation and provides
the population evolution of the excited state as
\begin{align}
P_{e}^{\mathrm{RWA}}(t)& =\frac{A^{2}}{4\Omega _{\mathrm{RWA}}^{2}}\sin
^{2}\left( \Omega _{\mathrm{RWA}}t/2\right)   \nonumber \\
& =-\frac{A^{2}}{16\Omega _{\mathrm{RWA}}^{2}}\left( e^{i\Omega _{\mathrm{RWA%
}}t}+e^{-i\Omega _{\mathrm{RWA}}t}+2\right) ,  \label{RWA}
\end{align}%
where $\Omega _{\mathrm{RWA}}=\sqrt{(\Delta -\omega )^{2}-(A/2)^{2}}$. The
population grows exponentially when $\Omega _{\mathrm{RWA}}$ is imaginary
and can exceed unity due to the non-conservation of probability in
non-Hermitian systems.

The  above quasi-eigenvalues in Eq.~ (\ref{quasienergies}) allow us to analytically
study the dynamical properties of the NHRM. Using the similarity transformation
(\ref{transfer}) and the unitary transformation $\hat{R}(t)$, we can define
the time evolution operator as
\begin{equation}
\hat{U}(t)=e^{-\hat{S}(t)}\hat{R}^{\dagger }(t)e^{-i\tilde{H}t}\hat{R}(0)e^{%
\hat{S}(0)},  \label{operator_time}
\end{equation}%
by which we can derive analytically the time-dependent wavefunction $|\psi
(t)\rangle $ starting from an initial state $|\psi _{i}\rangle $. The time
evolution operator in Eq.~ (\ref{operator_time}) applies a sequence of
transformations to the initial state, moving it to a new gauge. The state
evolves under the time-independent Hamiltonian $\tilde{H}$ until time $t$. Finally, we arrive at the evolved state to the original frame using
the transformation $e^{-\hat{S}(t)}\hat{R}^{\dagger }(t)$, yielding the
final state $|\psi (t)\rangle =\hat{U}(t)|\psi _{i}\rangle $.

The atom is initially prepared in the ground state, so $|\psi _{i}\rangle
=(0\ 1)^{\text{T}}$. In our analysis, we focus on the atomic population of
the excited state, denoted as $P_{e}(t)=|c_{+}(t)|^{2}$, representing the
probability of finding the atom in its excited state at time $t$. This
initial state has a definite  parity, so its evolution involves only  
eigenstates in the same parity subspace. The analytical result can be
written as:
\begin{widetext}
\begin{align}
P_{e}& =L^{2}\left[ \frac{\tilde{A}}{4\Omega _{R}}\sin (\Omega _{R}t)\right]
^{2}+LP\frac{\tilde{A}}{4\Omega _{R}}\left[ \sin (2\Omega _{R}t)\cos (\omega
t)-\frac{\tilde{\Delta}}{2\Omega _{R}}[\cos (2\Omega _{R}t)-1]\sin (\omega t)%
\right]   \notag \\
& +P^{2}\left[ \cos ^{2}(\Omega _{R}t)+\left( \frac{\tilde{\Delta}}{2\Omega
_{R}}\right) ^{2}\sin ^{2}(\Omega _{R}t)\right] ^{2},  \label{PE}
\end{align}%
\end{widetext}
where%
\begin{align*}
L& =I_{0}\left( \frac{A\alpha }{2\omega }\right) +2\sum_{n=1}^{\infty
}(-1)^{n}I_{2n}\left( \frac{A\alpha }{2\omega }\right) \cos (2n\omega t), \\
P& =2\sum_{n=0}^{\infty }(-1)^{(n+1)}I_{2n+1}\left( \frac{A\alpha }{2\omega }%
\right) \cos [(2n+1)\omega t]. \label{Evolution}
\end{align*}
Here,  $\Omega _{R}=\left( \varepsilon _{+}-\varepsilon
_{-}\right) /2$ is the Rabi frequency from Eq.~(\ref{quasienergies}). $%
2\Omega _{R}$ measures the gap between the two branches of the
quasi-eigenvalues for the same $n$ in the eigenvalue spectrum, as
illustrated in Fig.~\ref{fig:spectrum} (b) by the black double arrow
connecting the open circles. Adjacent quasi-eigenvalues with the same parity are shifted by $2\omega$. The quasi-eigenvalue differences dominating the dynamics can include $2\omega$, $2\omega \pm \Omega_{R}$, as well as other combinations and multiples of $2\omega$ and $2\Omega_{R}$.		

In Fig.~\ref{fig:dynamics}, we show the time evolution of the atom at three
representative points: A ($\Delta /\omega =2.5$, $A/\omega =1$), B ($\Delta
/\omega =3.5$, $A/\omega =3$), and C ($\Delta /\omega =2.5$, $A/\omega =4$%
). These points lie in different regions of the phase diagram in Fig.~\ref%
{fig:ptphase}, and we analyze them using the present analytical approach (%
\ref{PE}), numerical simulations, and the RWA (\ref{RWA}).  As illustrated in
Fig.~\ref{fig:dynamics} (c), at point C in the $\mathcal{PT}$-broken phase,
dominated by single-photon resonance, the population in the excited
state grows roughly exponentially, eventually surpassing unity. This feature
is fully described by Eq. ~(\ref{PE}). This intriguing emergent phenomenon
has significant implications for non-Hermitian dynamics. Unlike Hermitian
systems, our model equivalently treats open systems, leading to a breakdown
in probability conservation. Interestingly, the present analytical results
for population evolution perfectly agree with the numerical
results. The RWA can only accurately describe population evolution in the
short-time regime. While the RWA can predict the dominant exponential
behavior (cf. Eq.~(\ref{RWA})), it underestimates the growth rate.
Additionally, we find that the RWA yields a smoothed evolution without the
oscillatory feature  due to the neglect of counter-rotating terms.

\begin{figure}[t]
\includegraphics[width=\linewidth]{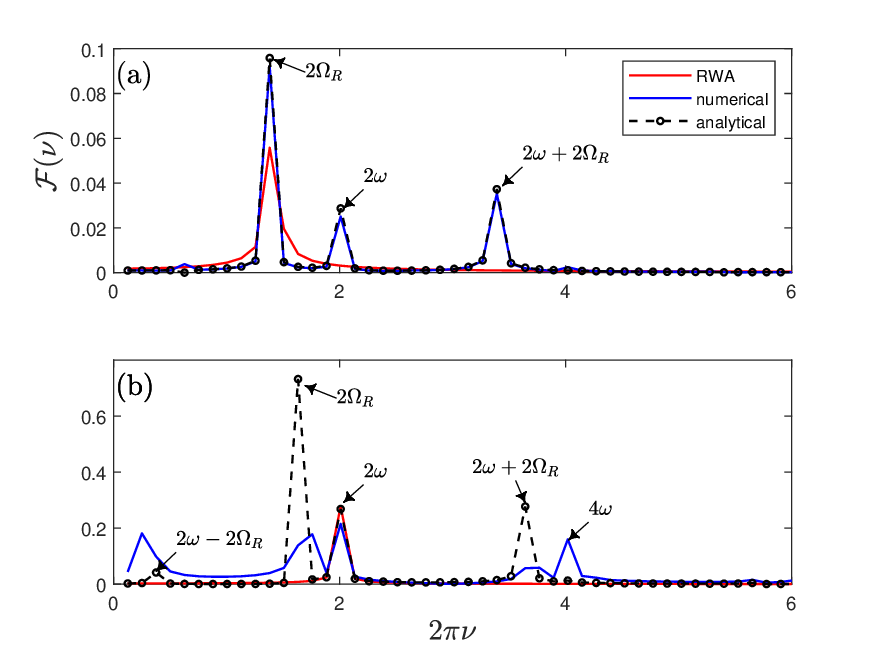}
\caption{ (a) and (b): The Fourier transform $\mathcal{F}(\nu ) $ of $P_e(t)$  in  Fig. ~\ref{fig:dynamics}  (a) and  (b), respectively. The open circles denote the analytical results, and the blue line represents the numerical results. The red line represents the results in  the  RWA.  All frequency peaks in the analytical results are labeled  with their corresponding expressions.  }
\label{fig:fs}
\end{figure}

In the $\mathcal{PT}$-unbroken phase, represented in Fig.~\ref{fig:dynamics}
(a) and (b), stable oscillations with multiple frequencies and varying
amplitudes are observed, in contrast to the $\mathcal{PT}$-broken phase,
where exponential dynamics dominate. The present analytical results, given
by Eq.~(\ref{PE}), perfectly describe these oscillations at point A for $%
\Delta /\omega <3$. At point B, the analytical results match the exact
dynamics only at the very early stage and  deviate significantly
throughout the rest of the process. Nevertheless, they qualitatively capture
the multi-oscillation feature, including  oscillations with multiple
frequencies. 

To analyze the dynamics and stable oscillations in the $\mathcal{PT}$%
-unbroken phase in greater  detail, we perform a Fourier spectral analysis of the
population time evolution. The Fourier transform, a fundamental technique
for frequency decomposition, is defined as follows:
\[
\mathcal{F}(\nu )=\int_{0}^{\infty }F(t)e^{-i2\pi \nu t}dt,
\]%
where $\mathcal{F}(\nu )$ represents the spectrum as a function of frequency
$\nu $, and $F(t)$ denotes the observed time evolution as a function of time
$t$.

The frequency spectrum of the atomic population time evolution is shown
in Fig.~\ref{fig:fs}. At the point A, as shown in Fig.~\ref{fig:fs} (a), the
numerical frequency spectrum exhibits three peaks, precisely corresponding
to $2\Omega _{R}$, $2\omega $, and $2\omega +2\Omega _{R}$, all these three
frequencies measure the gaps between the two quasi-eigenvalues within the
same parity in the analytical scheme. These frequencies also correspond to
the lowest three frequencies in  Eq.~(\ref{PE}), further demonstrating that our
analytical method accurately describes the time evolution in the $\mathcal{PT%
}$-unbroken phase for $\Delta /\omega <3$.

As shown in Fig.~\ref{fig:fs} (b) for point B, the exact frequency spectrum
exhibits five peaks, while our analytical results yield four main peaks,
missing only the last one. The analytical frequency peaks are marked with
arrows, indicating  the differences between two quasi-eigenvalues of the
same parity. In both Fig.~\ref{fig:fs} (a) and (b), the main peaks
corresponding to $2\Omega _{R}$, $2\omega $, and $2\omega +2\Omega _{R}$ also appear in the exact dynamics. In panel (b), however, the additional
peaks around $2\omega -2\Omega _{R}$ and $4\omega $ in the numerically exact
dynamics are only weakly captured in the analytical results. At point B,
the atomic frequency $\Delta/\omega=3.5 $ lies beyond the three-photon resonance
region,  so additional higher-order frequency terms are required. Therefore, to obtain a more precise description, additional similarity transformations should be added in Eq. ~(\ref{transfer}). For qualitative studies, the current analytical scheme, which uses a
single similarity transformation, can capture the main features of the
dynamics.

We have also extensively examine the dynamics for large atomic frequencies $\Delta/\omega \geq 4$, in the $\mathcal{PT}$-unbroken phase, as  demonstrated  in  Appendix B.  Our analytical results agree well with the exact ones as long as the coupling strength is not too strong. 
The RWA can only provide the primary oscillation frequency, known as the
Rabi frequency, and thus cannot fully capture the complex features of the
time evolution, as shown in Figs.~\ref{fig:dynamics}, ~\ref{fig:fs}, and~\ref{appendixB}. The RWA Rabi frequency also deviates from the exact values at strong coupling, whereas the analytical Rabi frequency agrees well with the exact
values in all cases.

\section{Bloch-Siegert shift}
In this section, we will discuss the Bloch-Siegert (BS) shift  in the
NHRM, which is a key issue in its Hermitian counterpart
~\cite{Shirley1965,Yan2015,Bloch1940}.
\begin{figure}[t]
\includegraphics[width=\linewidth]{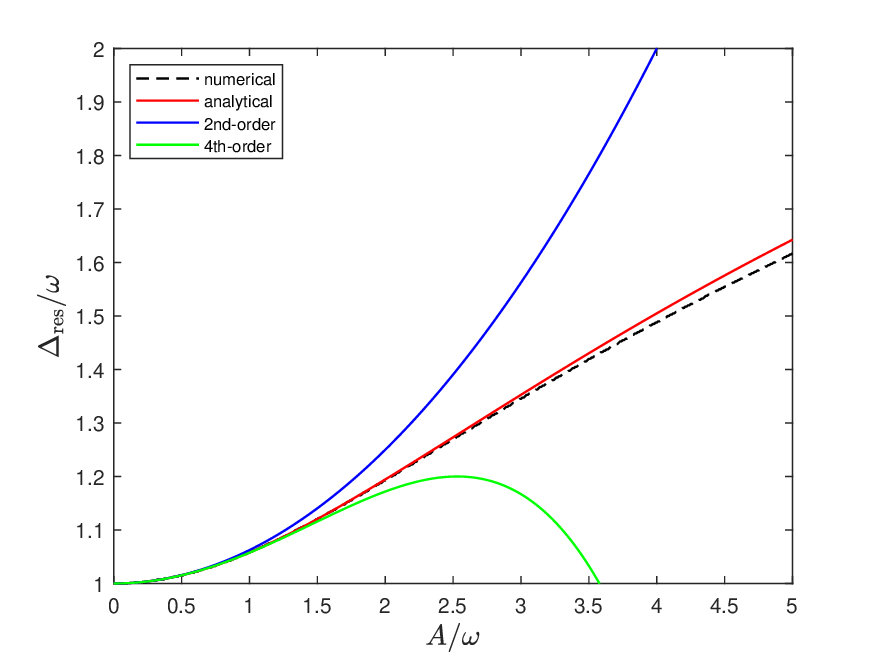}
\caption{The resonance frequency as a function of coupling strength $A/%
\protect\omega$ for numerical results (black dashed line) and the  
analytical results (red line) from Eq. ~(\ref{BS_shift}). The 2nd-order (blue) and 4th-order (green) approximations are also displayed.}
\label{fig:BSshift}
\end{figure}
In the Hermitian system, the BS shift represents the offset of the resonance
frequency $\Delta _{\mathrm{res}}$ from the original driven frequency, i.e.,  $%
\delta \omega _{\mathrm{BS}}=\Delta _{\mathrm{res}}-\omega$. The
resonance frequency is determined by the maximum time-averaged
transition probability, corresponding to the largest oscillation amplitude.
However, in the non-Hermitian system, when the frequency reaches resonance,
the probability diverges rapidly. In this case, the quasienergies exhibit
the largest imaginary component. The condition of the resonance is 
\begin{align}
& \frac{\partial \Omega _{R}^{2}}{\partial \Delta }=2\left[ \Delta
I_{0}\left( \frac{A}{\omega }\alpha \right) -\omega \right] \left[ I_{0}\left( \frac{A}{\omega }\alpha \right) \right.
\notag \\
& \left. + \frac{\Delta A}{\omega}I_{1}\left( \frac{A}{\omega }\alpha \right)\frac{\partial \alpha }{\partial \Delta }\right] +2A^{2}(1-\alpha )\frac{\partial \alpha }{\partial \Delta }=0,  \label{BS_shift}
\end{align}%
where  $\partial \alpha /\partial \omega $ can be obtained from Eq.~(\ref{transition}) as
\begin{equation*}
\frac{\partial \alpha }{\partial \Delta } = \frac{-2\omega I_{1}\left( \frac{A}{\omega } \alpha \right)  }{A \left\{ \omega +\Delta \left[ I_{0}\left(
\frac{A}{\omega }\alpha \right) +I_{2}\left( \frac{A}{\omega }\alpha \right) %
\right] \right\} }.
\end{equation*}%
Solving Eq. ~(\ref{BS_shift}) yields $\Delta_{\mathrm{res}}$, allowing us to compute the BS shift in the present analytical scheme.	

A simple analytical expression for the BS shift, expanded in powers of the coupling strength $A/\omega$, can also be derived.	 In this process, we  expand $I_0\left(\frac{A}{\omega}\alpha\right)$ and $I_1\left(\frac{A}{\omega}\alpha\right)$ up to the fourth order in $A$,	
\begin{align*}
    I_0\left(\frac{A}{\omega}\alpha\right) &= 1 + \frac{1}{4}\left(\frac{A}{\omega}\alpha\right)^2 + \frac{1}{64}\left(\frac{A}{\omega}\alpha\right)^4+\mathcal{O}(A^6),\\
    I_1\left(\frac{A}{\omega}\alpha\right) &= \frac{1}{2}\left(\frac{A}{\omega}\alpha\right) + \frac{1}{16}\left(\frac{A}{\omega}\alpha\right)^3+\mathcal{O}(A^5).
\end{align*}
Using Eq.~(\ref{transition}), we  expand the coefficient $\alpha$ up to second order in $A$,
\begin{align*}
    \alpha = \frac{\omega}{\Delta+\omega} \left[ 1 - \frac{\Delta}{8(\omega+\Delta)^3}A^2 + \mathcal{O}(A^4)\right].
\end{align*}
The minus sign in the second term arises from the imaginary coupling $iA$, which changes the sign in the Hermitian counterpart ~\cite{Lv2012}.	Therefore, the modulated Rabi frequency, up to fourth order in $A$, is given by	
\begin{align}
    \Omega_R^2 \simeq \left(\Delta - \omega\right)^2 - \frac{\Delta A^2}{2\left(\Delta + \omega\right)} - \frac{\Delta A^4}{32\left(\Delta + \omega\right)^3}.
\end{align}
We can now obtain the resonance frequency,
\begin{align}
    \Delta_{\rm res} = \omega + \frac{1}{16}\frac{A^2}{\omega} - \frac{5}{1024}\frac{A^4}{\omega^3} + \mathcal{O}(A^6).
\end{align}
Note that the $A^2$ term is consistent with the one found in  perturbation theory~\cite{Lee2015}.  Additionally, by applying the concise similarity transformation method, we can derive the higher-order solution for the BS shift.

The analytical BS shift, based on Eq. ~(\ref{BS_shift}),  is shown by the red line in Fig.~\ref{fig:BSshift} .  Interestingly, our results match  the numerical data very well across a wide range of coupling strengths, up to $A/\omega = 5$.	 Note that the  perturbation result for the BS in Ref. ~\cite{Lee2015},  marked in blue,  begins to deviate from the exact values at $A/\omega=1.5$. Surprisingly, the presence of the $\mathcal{PT}$-broken phase in this non-Hermitian model does not significantly affect the BS shift across a broad range of coupling strengths, even when passing through multi-photon resonance. This may be due to the fact that the BS shift can be expressed as a function of $(iA)^2$.		

\section{conclusion}

This work studies the NHRM analytically. First,  we derive a transformed NHRM in the RWA form using  a single similarity transformation. Next, we
obtain an effective Hamiltonian for the non-Hermitian two-level system by
applying a rotating operation, which explicitly yields two eigenvalues and
an EP. Thus, the quasi-eigenvalues can thus be obtained from the two eigenvalues
by shifting them by a multiple of the field frequency. The  exact
quasi-eigenvalues can be determined numerically using Floquet theory. Notably, our
analytical results for the quasi-eigenvalues agree excellently with the
numerical results in the single-photon resonance regime. The primary $%
\mathcal{PT}$-broken phase boundary is also detected in a wide parameter
regime,  extending even beyond the multi-photon resonance region.

We also uncover a Floquet parity symmetry that commutes with the Floquet
Hamiltonian of the NHRM. The conserved parity decomposes the entire Floquet
space into two subspaces with different parity. The quasi-eigenvalues
are  associated with one of two parities. Quasi-eigenvalues shifted
by even multiples of the field frequency, $2n\omega$,  belong to  the same
parity. When the atomic frequency exceeds an odd multiple of the field
frequency, i.e., $\Delta >(2n+1)\omega $, two adjacent quasi-eigenvalues
within the same parity will intersect. However, they cannot truly cross
without an additional symmetry. Therefore, a complex quasi-eigenvalue emerges
to resolve this issue, leading to the secondary $\mathcal{PT}$-broken phase.
Interestingly, in this region, our analytical real quasi-eigenvalues 
cross, predicting the $\mathcal{PT}$-broken location, though it 
cannot describe the $\mathcal{PT}$-broken.

The dynamics are also investigated within the analytical scheme. In the
single-photon resonance regime, i.e., $\Delta /\omega <3$, the time evolution of the
atomic population, starting from the atomic upper state, as predicted by the
analytical scheme, closely matches the numerical results. Even in the higher-photon resonance regime, Fourier transform analysis shows that the analytical scheme can captures  multi-oscillations, with the dominant  frequencies aligning with those observed in the exact dynamics. Additionally, the BS shift, as derived  analytically in the present scheme, agrees well with the exact result over a wide coupling regime. This further highlights that the  simple analytical scheme effectively  captures the main features of this non-Hermitian qubit-cavity coupling model.	

To accurately  describe the secondary $\mathcal{PT}$-broken phase and dynamics in higher-order resonance regimes, additional similarity transformations are required. The research along this direction is in progressing. The proposed analytical method  for the NHRM may offers a valuable alternative for exploring open systems and atom-field interactions. Its versatility extends to more complex models, including those with multiple cavity frequencies, higher-order interactions, or additional qubits. 

\begin{acknowledgments}
		 This work is supported by the National Key R$\&$D  Program of China (Grant No.  2024YFA1408900) and  the National Natural Science Foundation of China (Grant No.  12305032).
	\end{acknowledgments}

\addcontentsline{toc}{chapter}{Appendix A: Numerical method for quasi-energy
spectrum}

\section*{Appendix A: Numerical method for quasi-energy spectrum}

\begin{figure}[t]
\subfloat{\includegraphics[width=\linewidth]{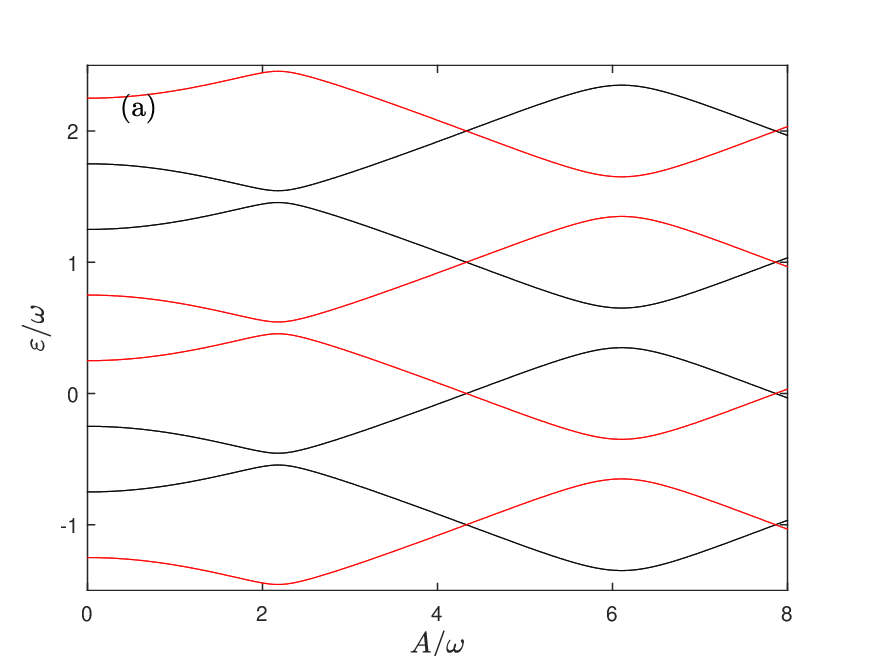}}\vspace{%
-2pt} \subfloat{\includegraphics[width=\linewidth]{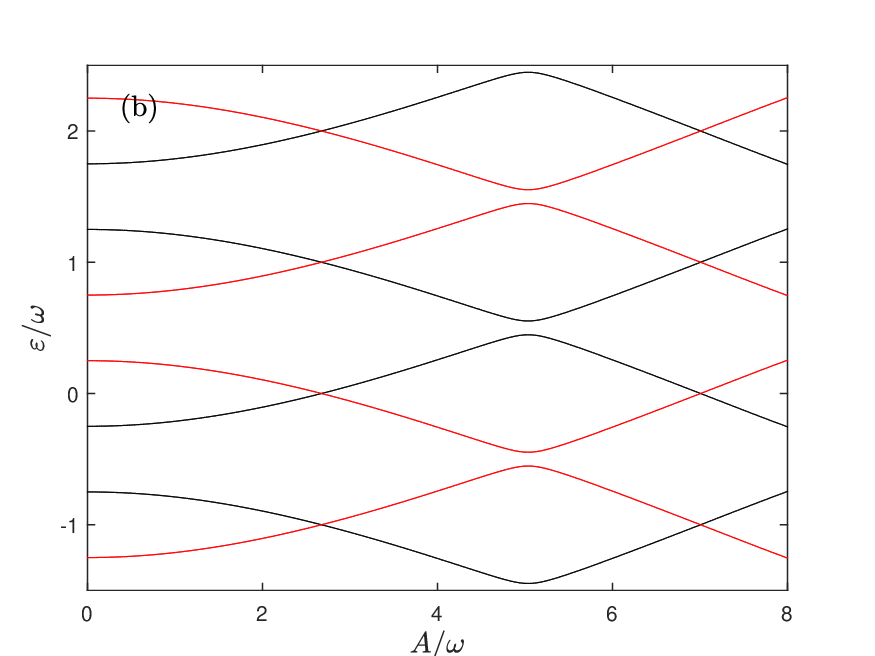}}
\caption{The quasi-eigenvalue spectrum, $\protect\varepsilon/\protect\omega$, as a function of the coupling strength, $A/\protect\omega$, in the Hermitian semiclassical Rabi model for  (a)  $\Delta/\protect\omega = 2.5$  and  (b) $\Delta/\protect\omega = 3.5$.  The black (red) dots represent the Floquet quasi-energies obtained by exact diagonalization of the Floquet Hamiltonian matrix in the odd (even) parity subspaces.	}
\label{Hermitianspectrum}
\end{figure}
\begin{figure}[thp]
\includegraphics[width=\linewidth]{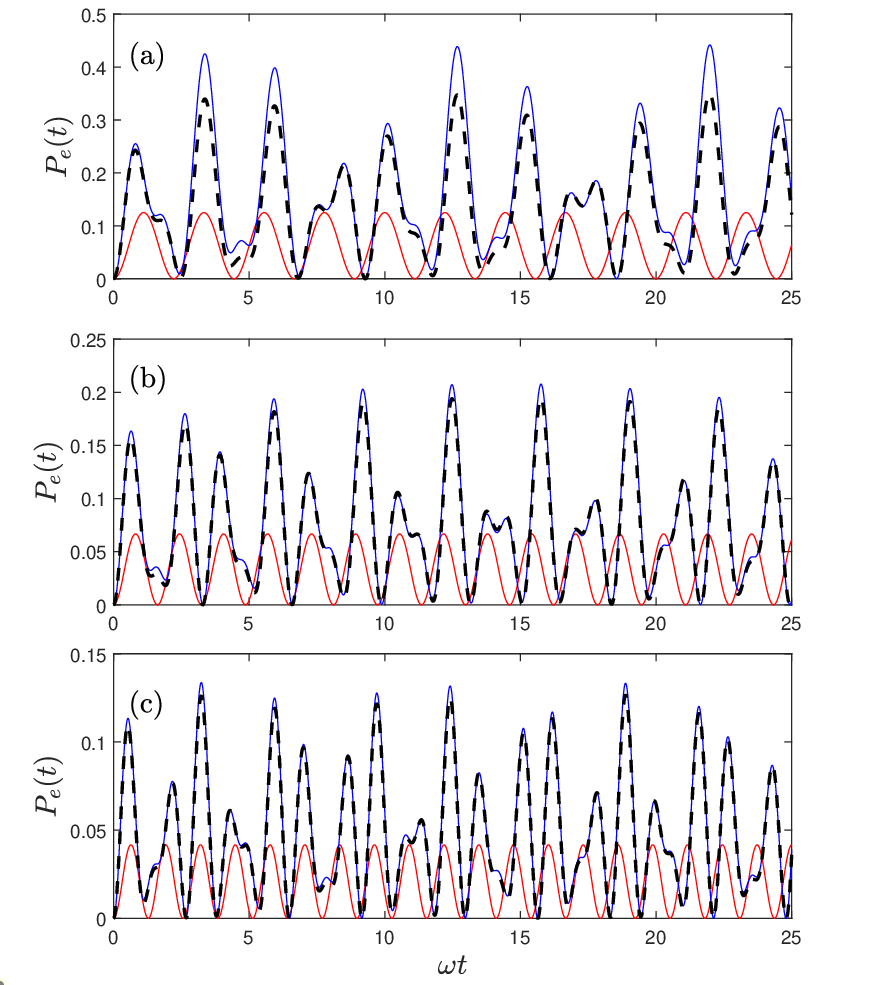}
\caption{Time evolution of the excited state population $P_{e}(t)$ at three  more points in the $\mathcal{PT}$-unbroken phase with different parameters, (a)$\Delta=4\protect\omega $, $A=2\omega $; (b)$\Delta=5\omega $, $A=2\omega $; and (c)$\Delta=6\omega $, $A=2\omega$. The analytical results are denoted by black dashed lines, and the numerical results are represented by the blue solid line. The red line denotes the results in rotating-wave approximation from Eq.~ (\ref{RWA}) .}\label{appendixB}
\end{figure}

In this Appendix, we provide a brief description of the Floquet picture for the NHRM. A complete set of orthonormal basis states $|\alpha ,m(t)\rangle \rangle $
of space $\mathscr{F}$ can be constructed by combining a complete set of
orthogonal and normalized basis states of $\hat{H}$, $|\alpha \rangle =|\pm
\rangle $, with the complete set of time-periodic functions $e^{im\omega t}$
labeled by the integer $m$,
\[
|\alpha ,m(t)\rangle \rangle =|\pm \rangle \mathrm{e}^{im\omega t}.
\]%
Using the definition of the scalar product in the extended Floquet Hilbert space, the matrix elements can be obtained $\mathcal{F}_{m^{\prime }m}^{\alpha
^{\prime }\alpha }=\langle \langle \alpha ^{\prime },m^{\prime }|\hat{%
\mathcal{F}}|\alpha ,m\rangle \rangle $ of the Floquet Hamiltonian $\hat{%
\mathcal{F}}=\hat{H}(t)-i\frac{\partial }{\partial t}$ with respect to the
basis $|\alpha ,m\rangle \rangle $,
\begin{align}
\mathcal{F}_{m^{\prime }m}^{\alpha ^{\prime }\alpha }& =\frac{1}{T}%
\int_{0}^{T}\mathrm{d}t\ \mathrm{e}^{-im^{\prime }\omega t}\langle \alpha
^{\prime }|\hat{H}(t)-i\frac{\partial }{\partial t}|\alpha \rangle \mathrm{e}%
^{im\omega t}  \nonumber \\
& =\langle \alpha ^{\prime }|H_{m^{\prime }-m}|\alpha \rangle +\delta
_{m^{\prime }m}\delta _{\alpha ^{\prime }\alpha }m\omega ,  \label{H_F}
\end{align}%
where
\[
H_{m^{\prime }-m}=\frac{1}{T}\int_{0}^{T}\mathrm{d}t\mathrm{e}^{-\mathrm{i}%
(m^{\prime }-m)\omega t}\hat{H}(t).
\]%
In the matrix form, the Floquet Hamiltonian can be expressed as
\begin{align}
    \begin{pmatrix} \ddots&\vdots&\vdots&\vdots&\vdots&\ddots\\
\cdots&\frac{\Delta}{2}-\omega&0&0&i\frac{A}{4}&\cdots\\
\cdots&0&-\frac{\Delta}{2}-\omega&i\frac{A}{4}&0&\cdots\\
\cdots&0&i\frac{A}{4}&\frac{\Delta}{2}&0&\cdots\\
\cdots&i\frac{A}{4}&0&0&-\frac{\Delta}{2}&\cdots\\
\ddots&\vdots&\vdots&\vdots&\vdots&\ddots \end{pmatrix}. \label{FH}
\end{align}
The quasi-eigenvalues are  the eigenvalues of this infinite matrix.	The exact numerical diagonalization of the matrix (\ref{FH}) is  performed using a large truncated space.

For comparison, we calculate the Floquet quasi-energies for the Hermitian counterpart using the same model parameters as in Fig.~\ref{fig:spectrum}. The results are presented in Fig.~\ref{Hermitianspectrum}. In the time-periodic Hermitian model, quasi-energies with the same parity  exhibit  only the avoided-crossing feature, in contrast to its non-Hermitian counterpart. Quasi-energies with different parities can cross. This is consistent with the Floquet parity symmetry.

\addcontentsline{toc}{chapter}{Appendix B: Dynamics of more parameters}

\section*{Appendix B: Dynamics of more parameters}

We can examine  the dynamics using  our analytic approach in $\mathcal{PT}$-unbroken phase over a broader parameter regime. As shown in Fig. \ref{appendixB}, we compare the dynamics for the different atomic frequencies: (a) $\Delta=4\omega$, (b) $\Delta=5\omega$, and (c) $\Delta=6\omega$, all at the same coupling strength $A=2\omega$. We observe that the population of the excited state maintains stable oscillations, and our analytical method matches  remarkably well  with the numerical results.

As shown in Fig.~\ref{appendixB} (c), even when the atomic frequency lies above the higher-order $\mathcal{PT}$-broken regions, such as the five-photon resonance region, our analytical approach still accurately captures  the dynamic behavior. This  suggests  that our analytic method is available for a wide range of coupling strengths and  arbitrary  atomic frequencies. Interestingly, as the frequency of atoms increases, the analytical results align more closely with  the numerical ones. 

\bibliographystyle{apsrev4-2}
\bibliography{nonhermisemi}

\end{document}